\documentclass[prl,amsmath,amssymb,twocolumn,floatfix,superscriptaddress]{revtex4}
\usepackage{graphicx}
\usepackage{times}

\begin{document}
\title{A Unified Approach to Nonlinear Transformation Materials}

\author{Sophia R. Sklan}
\affiliation{Department of Mechanical Engineering, University of Colorado Boulder,
Colorado 80309 USA}
\author{Baowen Li}
\affiliation{Department of Mechanical Engineering, University of Colorado Boulder,
Colorado 80309 USA}

\begin{abstract}
The advances in geometric approaches to optical devices due to transformation optics has led to the development of cloaks, concentrators, and other devices.
It has also been shown that transformation optics can be used to gravitational fields from general relativity.
However, the technique is currently constrained to linear devices, as a consistent approach to nonlinearity (including both the case of a nonlinear background medium and a nonlinear transformation) remains an open question.
Here we show that nonlinearity can be incorporated into transformation optics in a consistent way.
We use this to illustrate a number of novel effects, including cloaking an optical soliton, modeling nonlinear solutions to Einstein's field equations, controlling transport in a Debye solid, and developing a set of constitutive to relations for  relativistic cloaks in arbitrary nonlinear backgrounds.
\end{abstract}

%%% \date{\today}

\maketitle

\begin{figure}\begin{center}
\includegraphics[scale=0.35]{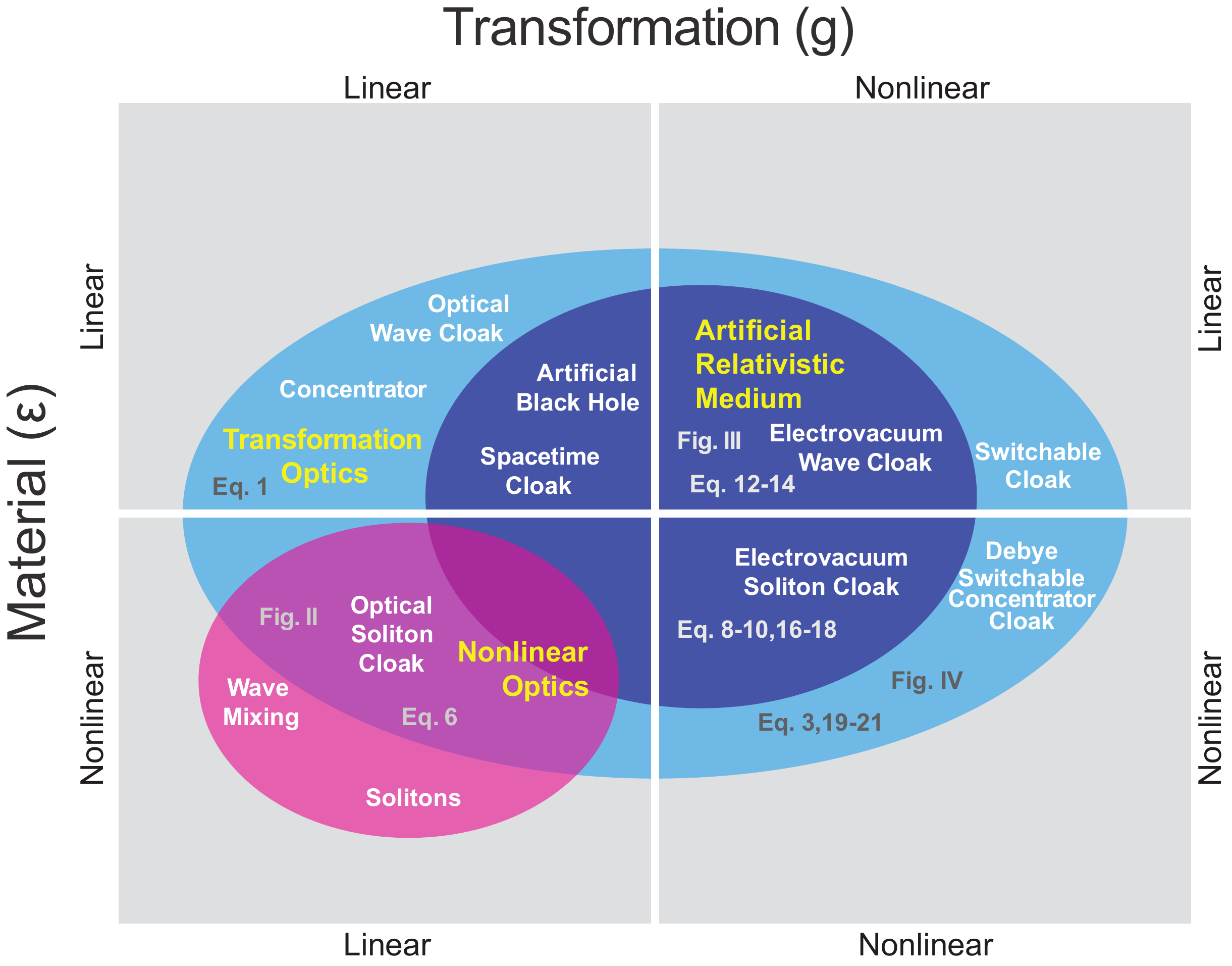}
\caption{\label{fig:model}Representation of our transformation optics framework. Background material ($\epsilon$) and coordinate transform ($g$) can be linear or nonlinear with respect to applied fields, making four mutually exclusive cases. Within this parameter space, certain combinations satisfy transformation optics requirements (light blue ellipse). A subset of these also satisfy Einstein's field equations (dark blue ellipse). When nonlinearity is included, effects from other fields, e.g. nonlinear optics (red circle) can become incorporated into transformation optics. Examples from each quadrant are labelled, with the nonlinear examples being explored in the text (except the switchabe cloak, discussed in Ref. \cite{NLTR}).}
\end{center}\end{figure}

Transformation optics \cite{TO1,TO2,TO3,TO4,TO5,EMCR1,EMCR2}, which uses geometric coordinate transformations derive the materials requirements of arbitrary devices, is a powerful technique.
Essentially, for any geometry there corresponds a material with identical transport.
With the correct geometry, it is possible to construct optical cloaks \cite{EMC1,RC1,CC1} and concentrators \cite{Conc} as well as analogues of these devices for other waves \cite{TA1,TA2,TA3,TA4,TA5} and even for diffusion \cite{TD1,TD2,TD3,TD4,TD5,TD6,TD7,NEMR}.
While many interpretations and formalisms of transformation optics exist, such as Jacobian transformations \cite{TO2}, scattering matrices \cite{PC1,PC2,PC3,PC4,PC5}, and conformal mappings \cite{TO1}, one of the most theoretically powerful interpretations comes from the metric formalism \cite{TOGR}.
All of these approaches agree that materials define an effective geometry, however the metric formalism is important since it allows us to further interpret the geometry.
In particular, certain geometries correspond to solutions to Einstein's field equations, which relate geometric curvature to gravitational forces.
Materials that mimic these geometries, or artificial relativistic media, constitute a subset of transformation optics materials (dark blue circle, Fig. \ref{fig:model})
that can effectively model relativistic effects \cite{TOGR}, such as black holes \cite{GRC1,GRC2} and gravitational lensing \cite{GRC3} or create novel devices such as the space-time cloak (which hides events instead of objects) \cite{STC}.

One limitation of transformation optics, however, is the necessity of using linear materials (materials whose properties do not change with electric field, pressure, temperature, etc.).
At present, the transformations that have been derived have exclusively been applied to linear media.
That is, the focus has been upon media equivalent to an isotropic, homogeneous, linear background medium embedded in curvilinear coordinates.
However, there is no necessity to maintain the constraint of linearity.
In thermal transformations, researchers have already considered the case of temperature dependent transformations (which we shall generalize as ``nonlinear transformations''), and shown how they are equivalent to a thermally nonlinear material embedded within a linear background \cite{NLTR,NLTR2}.
However, considerations of background nonlinearity have thus far been absent.
Moreover, nonlinear transformations lack the intuitive physical interpretation of linear transformation materials, where transport follows stationary geodesics.
This intuition is useful when developing devices where geodesics bend and shift with the applied field.

In this paper, we shall present a unified theory of nonlinear transformation optics.
We will consider both the case of a nonlinear background medium (bottom half of Fig. \ref{fig:model}) and nonlinear transformations (right half of Fig. \ref{fig:model}) in arbitrary combination.
We shall begin by generalizing transformation optics theory to incorporate these nonlinearities, then consider examples illustrating this formalism from each of the new, nonlinear quadrants of Fig. \ref{fig:model}.
Examples will be selected for their practical significance, physical intuition, and clarity.

%More persuasively, it is already known in general relativity that sufficiently intense electromagnetic fields in a vacuum will create nonlinear effects due to the gravitational field produced by their energy \cite{GR}.
%Thus, if the relativistic interpretation of transformation optics as effectively gravitational fields from fictitious matter distributions is to be taken as rigorous, a theory of nonlinear transformation optics must also exist.

\textbf{Nonlinear Transformation Formalism}: To begin, in linear transformation optics, the constitutive relation is \cite{TOGR}
\begin{equation}\label{eq:TOlin}
\epsilon^{ij}/\epsilon_0=\mu^{ij}/\mu_0=\frac{\sqrt{-g}}{\sqrt{\gamma}}\frac{g^{ij}}{g^{00}}
\end{equation}
where $g^{ij}$ is the metric in transformed coordinates, $g$ is the determinant of the metric, $g^{00}$ is the time-like component of the metric ($-1$ for a static transform) and $\gamma$ is the determinant of the untransformed metric (1 for Euclidean coordinates, $r^2$ for cylindrical, etc.).
That is,
\begin{equation}
g^{i^\prime j^\prime}=\frac{\partial x^{i^\prime}}{\partial x^i}\frac{\partial x^{j^\prime}}{\partial x^j} \gamma^{ij}
\end{equation}
where we have used Einstein summation notation for curvilinear coordinates (indices repeated as both subscript and superscript (covariant and contravariant) are summed, Latin indices are only over spatial dimensions, Greek indices are over space and time ($0^{th}$) dimensions).
Eq. \ref{eq:TOlin} can easily be generalized to a nonlinear transformation of a nonlinear background by the relation
\begin{equation}\label{eq:TOnlin}
\epsilon^{ij}(E)/\epsilon(E)=\mu^{ij}(E)/\mu(E)=\frac{\sqrt{-g(E)}}{\sqrt{\gamma}}\frac{g^{ij}(E)}{g^{00}(E)}
\end{equation}
where we have assumed that the nonlinearity is solely a function of electric field $E(r,t)=\sqrt{E^iE_i}$.

Note that the functional forms of $\epsilon(E)$ and $g(E)$ are arbitrary.
Assuming the coordinate transformation $x^i\to x^{i\prime}$ leaves Maxwell's equations (or the corresponding equation of motion for other fields) unchanged, except for a change of variables (i.e. $\mathcal{L}[E(x),g_0,\epsilon(x,E(x)),x]=\mathcal{L}[E(x^\prime),g(x^\prime,E(x^\prime)),\epsilon_0(E(x^\prime)),x^\prime ]$ for operator $\mathcal{L}$ that defines $E$), then the introduction of nonlinearity preserves transformation optics techniques, as the coordinates only enter the nonlinearity through the field. 

\begin{figure}\begin{center}
\includegraphics[scale=0.4]{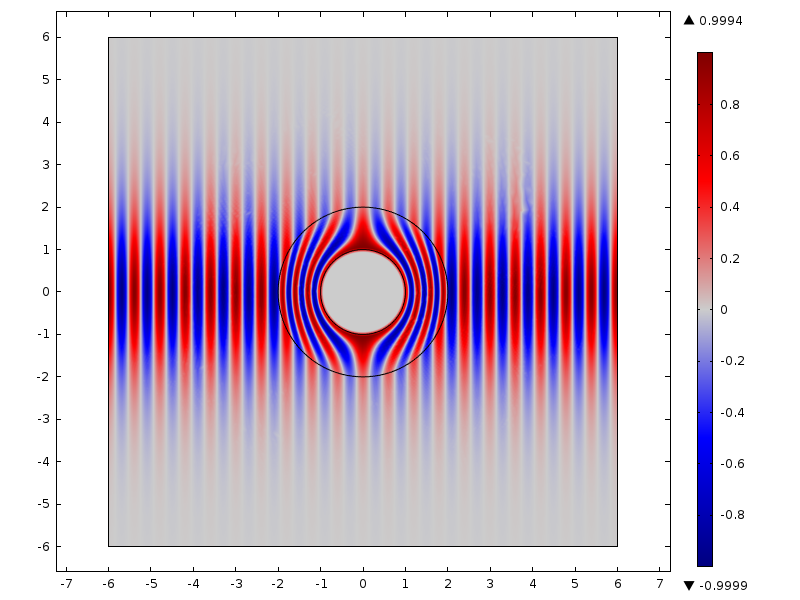}
\caption{\label{fig:soliton}Cloak of a medium with Kerr nonlinearity. Note the variation in wave amplitude, corresponding to first spatial soliton mode.}
\end{center}\end{figure}

\textbf{Nonlinear Background $-$ Linear Transform}: In particular, if the nonlinearity takes the form of a Kerr nonlinearity
\begin{equation}
P_i=\epsilon_{ij}E_j-\epsilon_{(0)}E_j=\epsilon_{(0)}(\chi^{(1)}_{ij}E_j+\chi^{(3)}_{ijkl}E_jE_kE_l)
\end{equation}
($P$ is polarization and $\chi$ is susceptibility, which we assume to be isotropic), Maxwell's equations remain unchanged under the cloaking transformation,
\begin{equation}\label{eq:TR}
r^\prime=a+\frac{b-a}{b}r.
\end{equation}
Thus, if we can find a solution to Maxwell's equations in Euclidean space with a Kerr nonlinearity, we can find a solution to Maxwell's equations with a Kerr cloak permittivity (lower left in Fig. \ref{fig:model})
\begin{equation}
\epsilon_{ij}=\epsilon_{(0)}(1+\chi^{(1)}+\chi^{(3)}E^2)\left[\begin{array}{ccc}
\frac{r-a}{r}&0&0\\
 0& \frac{r}{r-a}&0\\
0 &0  & (\frac{b}{b-a})^{2}\frac{r-a}{r}\end{array}\right]
\end{equation}
by writing the Euclidean solution in primed (i.e. cloak) coordinates.
The Kerr nonlinearity is a special case of nonlinear optics with an exactly solvable system for special values of intensity $E^2$ corresponding to optical soliton modes.
For concreteness, we select the first spatial soliton \cite{NLO},
\begin{equation}
\vec{E} =A_0\mathrm{sech}(y/y_0)e^{i(\omega t-k z+\gamma z)}\hat{y},
\end{equation}
where $A_0$ is the soliton intensity, $y_0=|A_0|\sqrt{3\chi^{(3)}}/2k$ is the pulse width, $k$ is the wave-vector, $\omega$ frequency, $\gamma=3k\chi^{(3)}|A_0|^2/4n_0^2$, and $n_0$ is the linear index of refraction.
The analytic solution is plotted in Fig. \ref{fig:soliton}. 
Note that the cloaking is exact in the analytic case, despite the nonlinear background.

\textbf{Linear Background $-$ Nonlinear Transform}: While we have seen that transformation optics is robust to background nonlinearity, that case is easier to understand.
The dynamics there are identical to nonlinear optics in Euclidean space, with the added linear transformation merely distorting the geodesics in fixed directions.
When the transformation is nonlinear, then the geodesics can change with changing intensity.
This makes, say, the combination of a nonlinear transform and the Kerr effect far harder to calculate.
Instead, we shall now consider only a nonlinear transform and fix the background to be linear (upper right in Fig. \ref{fig:model}).
We can apply physical intuition to the nonlinear transform by taking inspiration from the study of effective gravitational fields via linear transformation optics \cite{GRC1,GRC2}, where variations in the permittivity mimic the gravitational field produced by a mass distribution.
In that case, the metric used must satisfy Einstein's field equations
\begin{equation}\label{eq:Einstein}
G^{\mu\nu}=R^{\mu\nu}-\frac{1}{2}Rg^{\mu\nu}=\frac{8\pi G}{c^4}T^{\mu\nu}
\end{equation}
where $R_{\mu\nu}$ is the Ricci curvature tensor
\begin{gather}
R_{\mu\nu}=2\Gamma^\lambda_{\mu[\nu,\lambda]}+2\Gamma^\lambda_{\rho[\lambda}\Gamma^\rho_{\nu]\mu}\\
%\end{equation}
%($\Gamma^\mu_{\nu\lambda}$ is the Christoffel symbol
%\begin{equation}
\Gamma^\mu_{\nu\lambda}=\frac{1}{2}g^{\mu\rho}\left(g_{\nu\rho,\lambda}+g_{\lambda\rho,\nu}-g_{\nu\lambda,\rho}\right),
\end{gather}
(where $A_{\mu\nu,\rho}\equiv \frac{\partial A_{\mu\nu}}{\partial x^\rho}$, and $A_{[\mu\nu]}\equiv(A_{\mu\nu}-A_{\nu\mu})/2$), $R$ is the Ricci curvature scalar $R^\mu_\mu$, $G$ is the gravitational constant, $c$ is the speed of light, and $T^{\mu\nu}$ is the stress-energy tensor.
That is, a matter distribution is used to define $T^{\mu\nu}$, which then defines $g_{\mu\nu}$ via Eq. \ref{eq:Einstein}, thereby defining the equivalent $\epsilon,\mu$ via Eq. \ref{eq:TOlin}.
However, relativity also predicts that energy and mass are equivalent (as in the famous $\mathcal{E}=mc^2$).
As such, energy distributions can also define a stress-energy tensor and thereby 
produce a gravitational field \cite{GR}.

If the only source of energy is the electromagnetic field, then solutions to Eq. \ref{eq:Einstein} are referred to as electrovacuum solutions.
A material satisfying Eq. \ref{eq:TOnlin} with a metric transform obeying Eq. \ref{eq:Einstein}, then, will have a nonlinearity equivalent to the gravitational field produced by the electromagnetic field.

For a purely electromagnetic source, $T^{\mu\nu}$ is
\begin{equation}
T^{\mu\nu}=\left[\begin{array}{cc}
U & S_{i}/c\\
S_{i}/c & \sigma_{ij}\end{array}\right]
\end{equation}
where $U$ is the energy density
%\begin{equation}
%U=\frac{1}{2}\left(\epsilon E^2+\mu H^2\right),
%\end{equation}
$U=(\epsilon E^2+\mu H^2)/2$,
$\vec{S}$ is the Poynting vector $\vec{E}\times \vec{H}$, and $\sigma$ is the Maxwell stress tensor
%\begin{equation}
$\sigma_{ij}=\epsilon E_iE_j+\mu B_iB_j-\frac{1}{2}\left(\epsilon E^2+\mu H^2\right)\delta_{ij}.$
%\end{equation}
Notably, for a purely electromagnetic source, $T^\mu_\mu=0$ so $R=0$ and our equations simplify.

\begin{figure}\begin{center}
\includegraphics[scale=0.2]{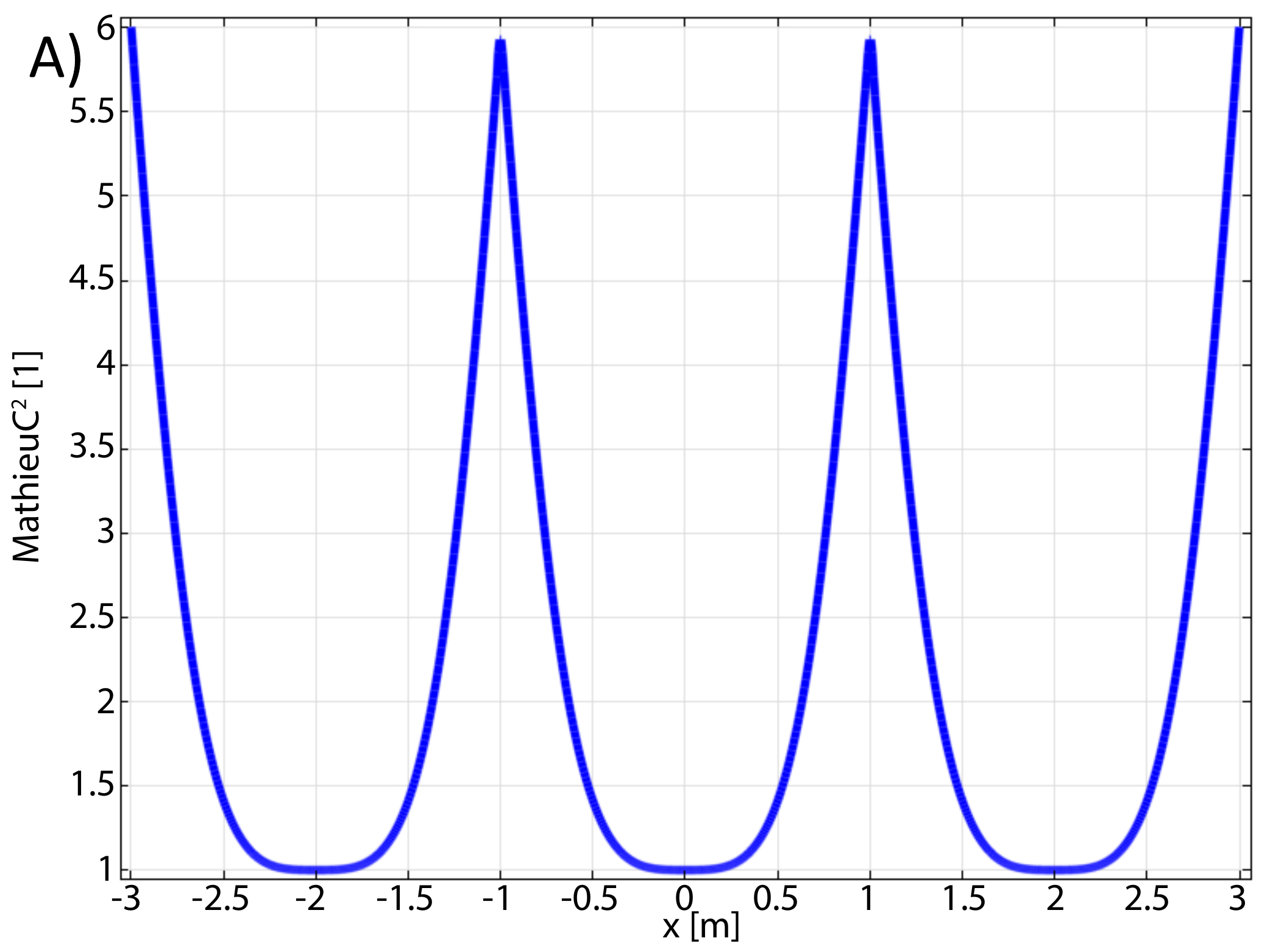}
\includegraphics[scale=0.2]{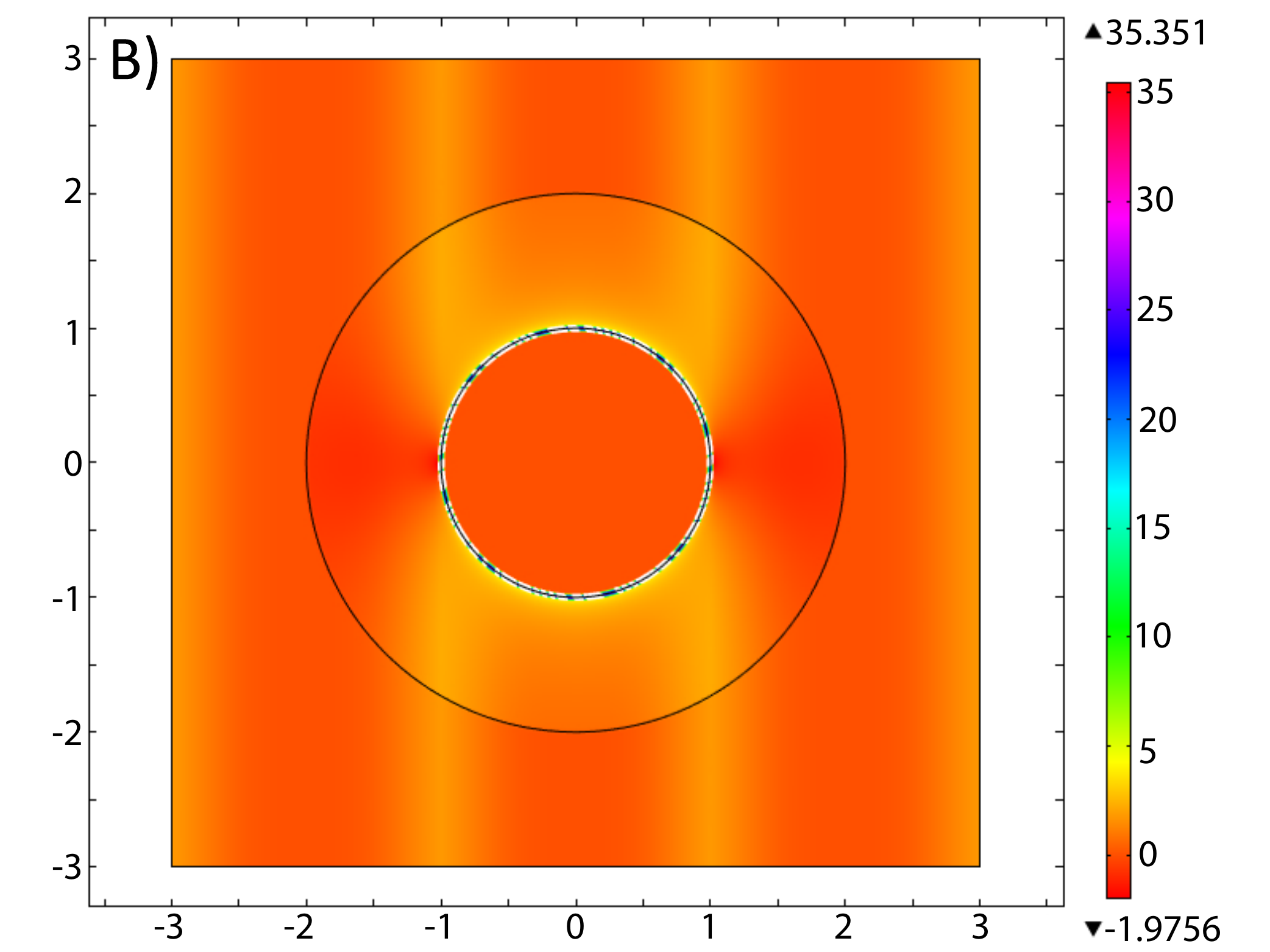}
\includegraphics[scale=0.2]{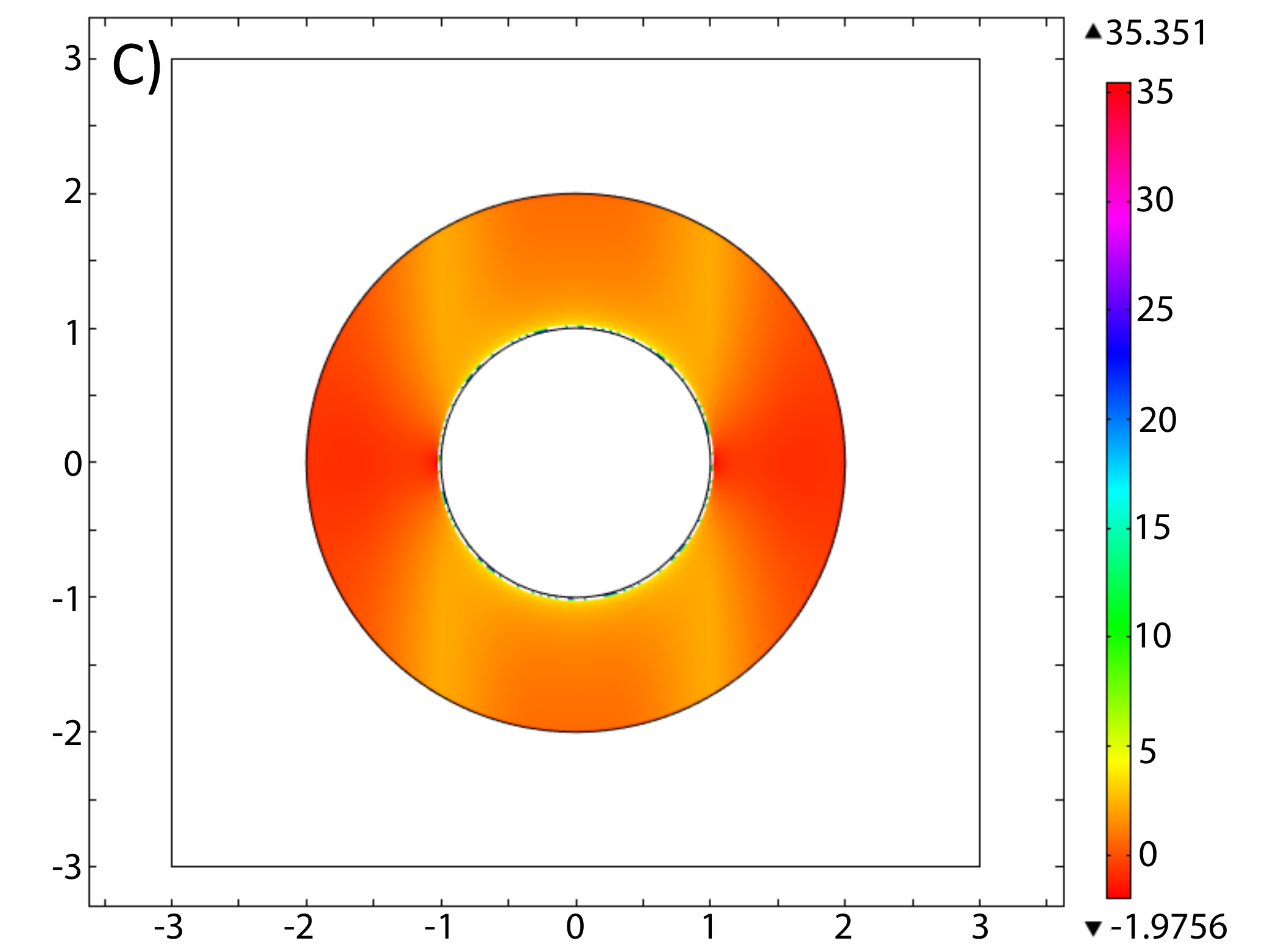}
\includegraphics[scale=0.2]{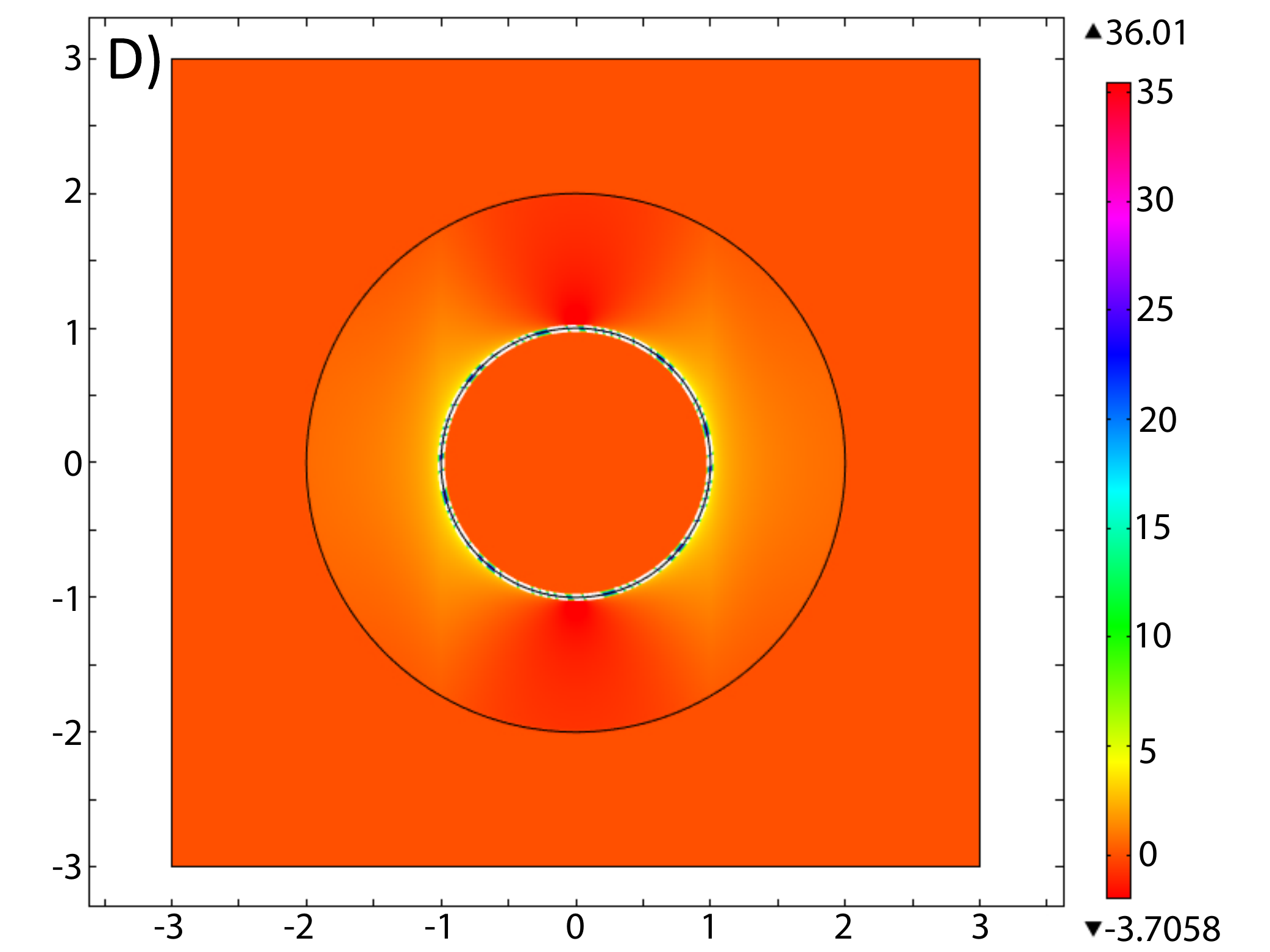}
\caption{\label{fig:GRC}Electrovacuum cloak solution for a linear background. (a) Functional dependence of the metric vs position at constant field strength, using a periodic continuation to preserve a finite metric. (b) Corresponding value of $\epsilon_{xx}$, plotted on a log scale to handle singularity at $r=a$. (c) Log scale of $\epsilon_{xy}$, which is only non-zero within the cloak. (d) Log scale of $\epsilon_{yy}$.}
\end{center}\end{figure}

To be more specific, we consider a plane wave solution $\vec{E}=|E|\cos(\omega t-\omega x/c)\hat{y}$.
If our background is linear, then $U(E)=\epsilon_0E^2$ (note that this is not the averaged energy, it retains space and time dependence), $\vec{S}=cU(E)\hat{x}$, and $\sigma_{ij}=-U(E)\delta^x_i\delta^x_j$.
We can then assume a metric of the form $g_{\mu\nu}\equiv$diag$[-1,1,f(ct-x),f(ct-x)]$ in Minkowski coordinates and use Eq. \ref{eq:Einstein} to get
\begin{equation}\label{eq:diffEQ}
U(E)=\frac{f^{\prime\prime}}{f}-\frac{1}{2}\left(\frac{f^{\prime}}{f}\right)^2=2h^{\prime\prime}/h,
\end{equation}
defining $h=f^2$.
Using the identity $2\cos^2(\phi)=1+\cos(2\phi)$, the stability condition $h\to1$ as $|E|\to0$, and the rotational symmetry (implying $-|E|$ should give the same solution as $|E|$), gives
\begin{equation}
f=\mathrm{MathieuC}^2\left(-\frac{|E|^2}{4E^2_0},\frac{|E|^2}{8E^2_0},\omega(ct-x)\right),
\end{equation}
where MathieuC is the Mathieu cosine function (which, because the first term is negative, behaves closer to $\cosh$ than cosine) and $E_0=\omega c/\sqrt{4\pi G\epsilon^{(0)}}$ is the natural electric field scale.
Note that $G$ only occurs in $E_0$, and so an effective gravitational effect can be tuned by changing $E_0$.
In Fig. \ref{fig:GRC}a, we plot $f$ , where we've exploited the periodicity of Eq. \ref{eq:diffEQ} to create a periodic continuation of $f$ (using the unmodified form results in an exponential growth of $\epsilon$).
We now consider the composite transform $T_{GR,C}=T_CT_{GR}$, to create a cloaked region within this artificial relativistic medium.
Using $f$, and Eqs. \ref{eq:TOnlin},\ref{eq:TR} we calculate the components $\epsilon_{xx},\epsilon_{xy},\epsilon_{yy}$ and plot them in Fig. \ref{fig:GRC}b-d.
Notably, we do not plot $E$ for this setup, as it is indistinguishable from the solution to a purely linear cloak.
This is expected, given that the form of $E$ was assumed in solving for $g$, but we can also show that Maxwell's equations reduce to
\begin{gather}
\partial_{tt}fE_x=\partial_{yy}E_x+\partial_{zz}E_x \nonumber\\
\partial_{tt}E_y=\partial_{xx}E_y+\partial_{z}f^{-1}\partial_{z}E_y\\
\partial_{tt}E_z=\partial_{xx}E_z+\partial_{y}f^{-1}\partial_{y}E_y,\nonumber
\end{gather}
($c\equiv1$) which remain unchanged from linear Euclidean background from waves transverse to $\hat{x}$.
A weaker test field, however, could detect the presence of the effective gravitational field if it propagated transversely to this electric field.

\textbf{Nonlinear Background $-$ Nonlinear Transform}: While the Mathieu cosine form of the nonlinear transform is helpful for illustrating the physical relevance of a nonlinear transform, an alternative formulation is preferable for developing materials prescriptions.
In particular, it is preferable in nonlinear optics to know the dependence of the susceptibility as a Taylor series in $E$
\begin{equation}
P=\epsilon_0\sum_{n=1}\chi^{(n)}E^n\equiv\sum_{n=0}(\epsilon_{(n)}-\epsilon_0\delta_n^0)E^{n+1}
\end{equation}
to some finite order.
In considering this problem, we shall once allow $\epsilon(E)$ to have an arbitrary nonlinearity, as that is most useful for design (bottom right of Fig. \ref{fig:model}).
Eq. \ref{eq:diffEQ} remains unchanged, save for a modification of $U(E)$ to reflect the new value of $\epsilon(E)$, but the solutions can no longer be expressed in terms of analytic functions.
Instead, we employ the Liouville-Neumann series technique to solve for $h(u)$, where $u=\omega(t-x/c)$.
That is, we consider a series expansion $h=\Sigma|E|^{2n}h_{n}(u)$, where $h^\prime_{n}(0)=0$, $h_{n}(0)=\delta_n^0=h_{0}$, and
\begin{equation}
h_{n+1}(u)=\int_{0}^{u}dw\int_{0}^{w}dvU(E(v))h_{n}(v)/2.
\end{equation}
We truncate our solution $f=h^2$ at $4^{th}$ order in $E$, as terms of that order and below are most relevant to nonlinear optics.
However, truncation means that our solution takes the form $f(E=|E|\sin(u),u)$, as some terms have a more depend upon $u^m$ (i.e. secular terms from nonlinear resonance) that cannot be factored without higher order terms (these likely correspond to the \textit{cosh} dependence in the Mathieu solution).

This solution $f$ gives the vacuum nonlinearity in a flat space-time.
We now apply the cloaking transformation Eq. \ref{eq:TR} in cylindrical coordinate to this metric and use Eq. \ref{eq:TOnlin} to get
\begin{gather}
\epsilon_{ij}/\epsilon(E)=\left[\begin{array}{ccc}
\frac{r-a}{r}&0&0\\
 0& \frac{r}{r-a}&0\\
0 &0  & (\frac{b}{b-a})^{2}\frac{r-a}{r}\end{array}\right]\nonumber\\
+\left(f(E,u)-1\right)\left[\begin{array}{ccc}
\frac{r-a}{r}\cos^{2}\theta & -\cos\theta\sin\theta&0\\
-\cos\theta\sin\theta & \frac{r}{r-a}\sin^{2}\theta&0\\
\\0&0&0\end{array}\right]\label{eq:epsTr}
\end{gather}
in transformed cylindrical coordinates ($\mu_{ij}/\mu(E)$ defined identically).
To verify that this leaves Maxwell's equations unchanged, it suffices to observe that this can also be written $\epsilon_{ij}(r,\theta,E)=\epsilon_{ij}^{(clk)}(r)\tilde{\epsilon}_{ij}^{(nl)}(\theta,E)$ (the nonlinear resonance should be modified to functions of $x^\prime$ as they derive from $E(r)$), so $\epsilon^{(clk)}$ will reproduce Eq. \ref{eq:TR}, transforming Maxwell's equations from a set of operators $L[r,\epsilon_{ij}^{(clk)}(r)\tilde{\epsilon}_{ij}^{(nl)}(\theta,E(r,\theta)),E(r,\theta)]$ to $L[r^\prime,\tilde{\epsilon}_{ij}^{(nl)}(\theta,E(r^\prime,\theta)),E(r^\prime,\theta)]$ as desired for a nonlinear cloaking transformation.
Notice that the first term in Eq. \ref{eq:epsTr} is the standard linear cloak (recall that $f(E=0)=1$ and the second is purely due to the vacuum nonlinearity.
We can thus define $\epsilon_{ij}(E)/\epsilon(E)\equiv\epsilon_{ij}^{(l)}+(f(E,u)-1) \epsilon_{ij}^{(nl)}$ for the linear and nonlinear coefficient matrices of Eq. \ref{eq:epsTr}.
Multiplying by $\epsilon(E)$ and Taylor expanding in $E$ thus gives
\begin{widetext}
\begin{gather}
\epsilon_{ij}(E)=\epsilon_{ij}^{(l)}\epsilon_{(0)}+\epsilon_{ij}^{(l)}\epsilon_{(1)}E+\left[\epsilon_{ij}^{(l)}\epsilon_{(2)}E^{2}+\epsilon_{ij}^{(nl)}\epsilon_{(0)}\frac{|E|^{2}u^{2}-E^{2}}{2E_0^2}\right] %\nonumber\\
+\left[\epsilon_{ij}^{(nl)}\epsilon_{(1)}\frac{24|E|^{3}u+9|E|^{2}Eu^{2}-24|E|^{2}E-13E^{3}}{18E_0^2}\right.\nonumber\\
+\left.\epsilon_{ij}^{(l)}\epsilon_{(3)}E^{3}\right]
+\left[\epsilon_{ij}^{(l)}\epsilon_{(4)}E^{4}+\epsilon_{ij}^{(nl)}\left(\epsilon_{(2)}\frac{3|E|^{4}u^{2}-5|E|^{2}E^{2}+4|E|^{2}E^{2}u^{2}-2E^{4}}{8E_0^2}+\frac{\epsilon^2_{(1)}}{\epsilon_{(0)}}\frac{12u|E|^3E-12|E|^2E^2-2E^4}{9E_0^2}\right)\right.\nonumber\\
+\left.\epsilon_{ij}^{(nl)}\left(\epsilon_{(0)}\frac{5|E|^{4}u^{4}+3|E|^{4}u^{2}+3\left(11-8u-6u^{2}\right)|E|^{2}E^{2}-24u|E|^{3}E\sqrt{|E|^{2}-E^{2}}-3E^{4}}{48E_0^4}\right)\right]+O(E^5). \label{eq:const}
\end{gather}
\end{widetext}
Notice that, although $E_0$ was originally defined in terms of the constant $G$, it is the only place that such constant enters into the transformed material equation.
Thus, we are free to redefine $E_0$ as any effective scale for the electric field strength, rather than the scale prescribed by Eq. \ref{eq:Einstein}.
That is, we can use transformation optics to model a nonlinear gravitational field with arbitrary strength $E_0(G_{eff})$.

\textbf{Transformation Media Extension}: Before considering our final example, it is worth stepping back and considering how these nonlinear transformation optics techniques could be extended to other forms of transformation media.
Acoustics is by far the easiest generalization, as there are straightforward mappings from transformation optics to transformation acoustics \cite{TA4}.
Heat transport and diffusion are more difficult, however.
While the introduction of field dependence to the already established thermal transformation \cite{TD1} holds $-$ i.e. that
\begin{gather}
\kappa^{ij}/\kappa_0(T)=g^{ij}(T)\\
\rho c_p/\rho_0(T)c_{p0}(T)=\sqrt{-g(T)}
\end{gather}
$-$ is valid, the diffusion equation is not Lorentz invariant and therefore is not a valid equation for the relativistic interpretation.
Thus, while transformation materials is applicable to nonlinear heat transport, it cannot be interpreted in terms of effective gravitational fields.
However, because transformation diffusion is defined for an isotropic background $\kappa_0$, with all anisotropy arising from the transformation, a further interpretation is plausible.
Both the background nonlinearity and isotropic nonlinear transform control the speed of diffusion at a given temperature, whereas the anisotropic aspect controls the preferential direction of diffusion as a function of temperature.

\begin{figure}\begin{center}
\includegraphics[scale=0.2]{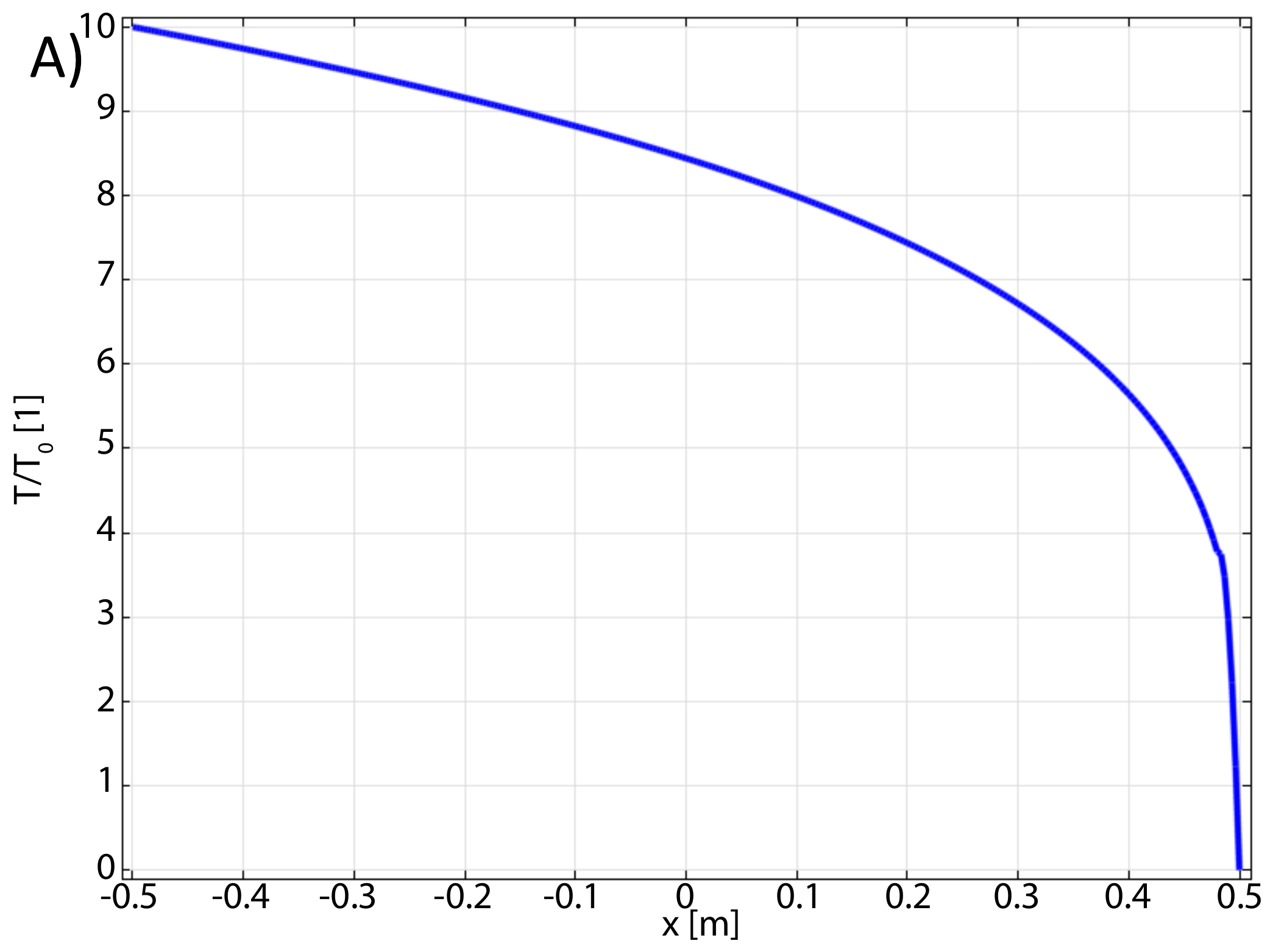}
\includegraphics[scale=0.2]{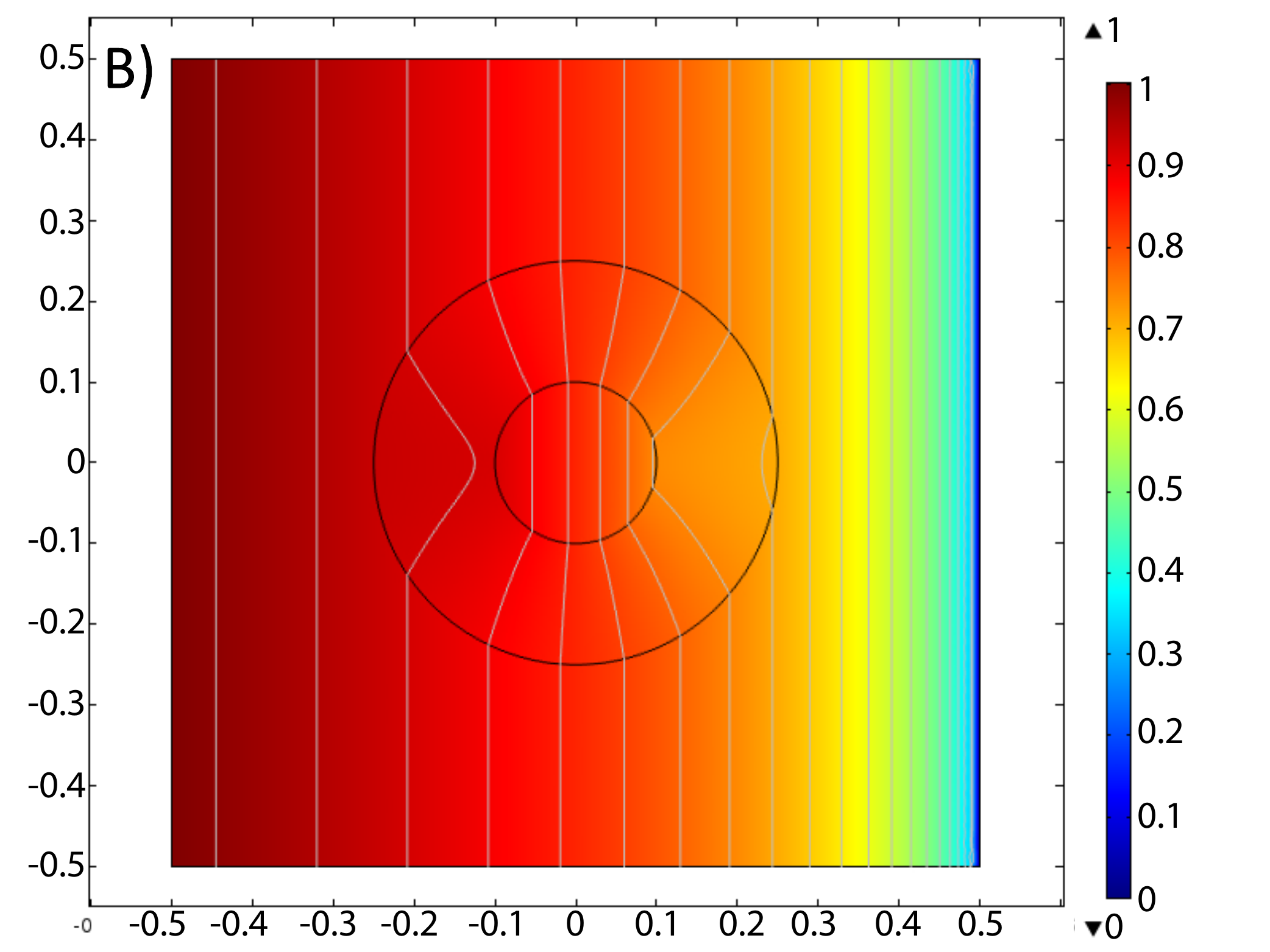}
\includegraphics[scale=0.2]{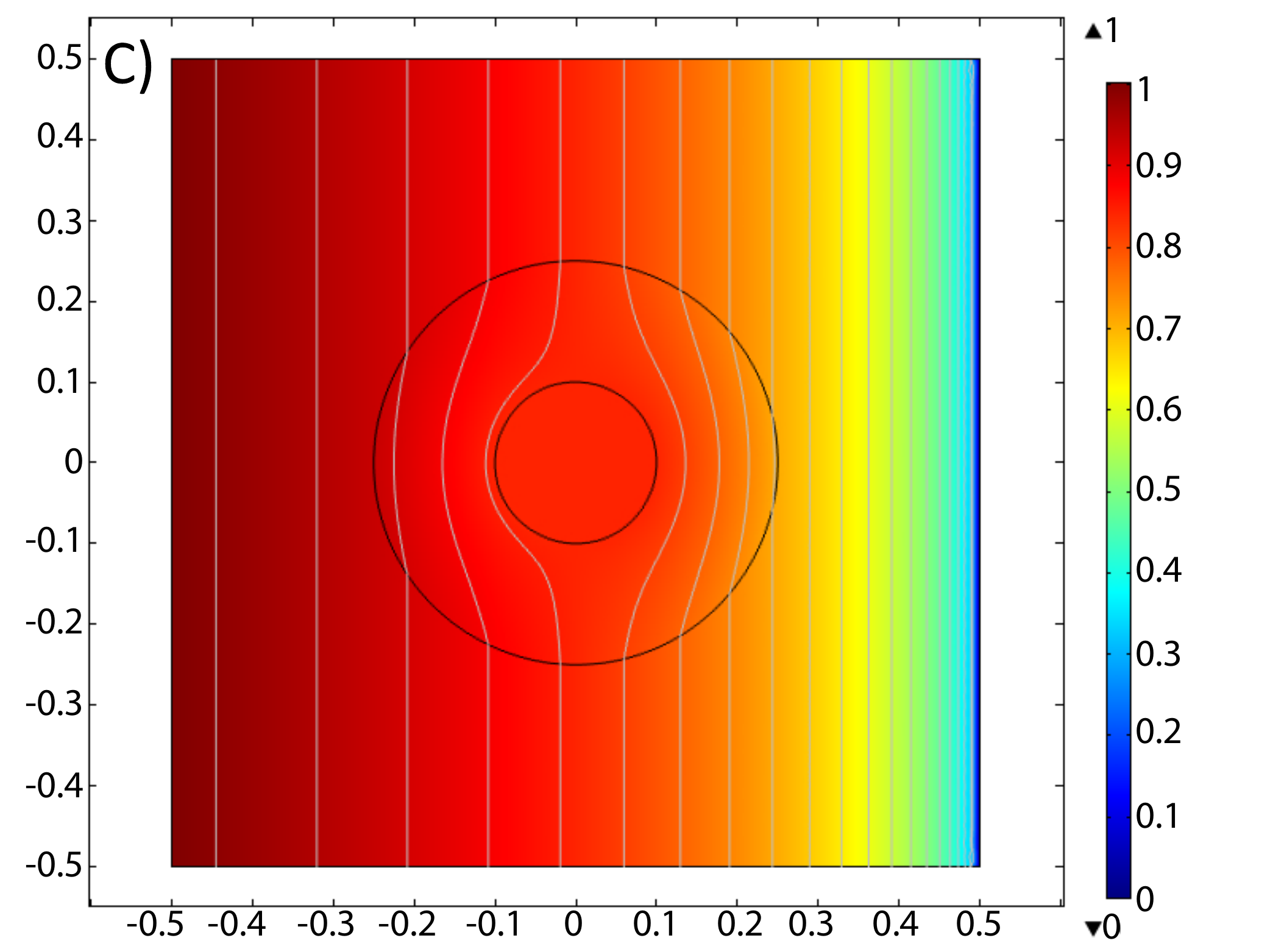}
\includegraphics[scale=0.2]{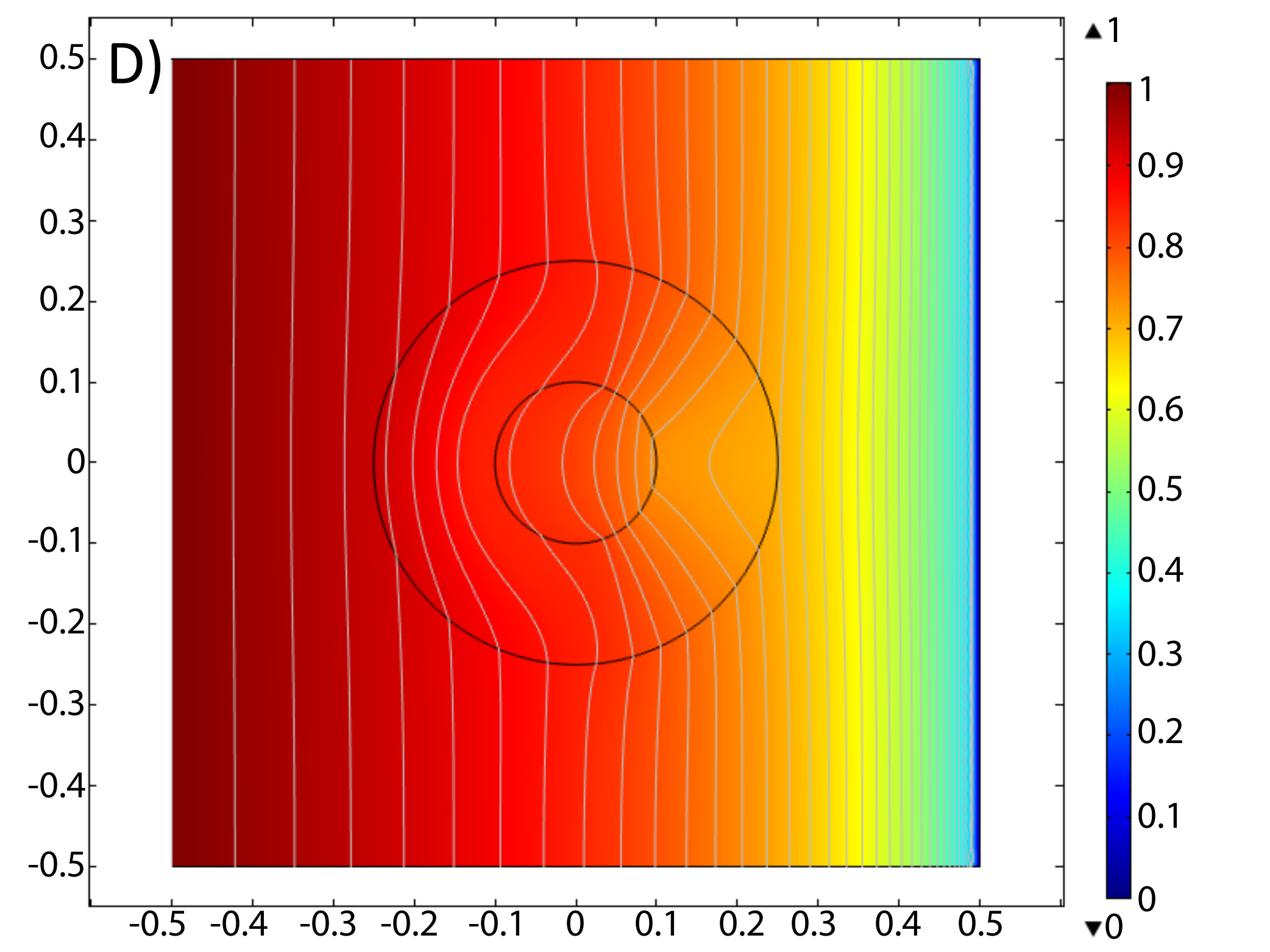}
\caption{\label{fig:Therm}Transformation diffusion of a Debye solid with switchable nonlinear concentrator/cloak transform. Steady state plots with $T(x=-L/2)=T_1,T(x=L/2)=0$. Note that isotherms (grey lines) are not evenly spaced due to Debye nonlinearity. (a) Far field temperature distribution. (b) Low temperature, $T_1=0.1T_0$, concentrator. (c) High temperature, $T_1=10T_0$, cloak. (d) Transitional, half cloak, half concentrator.}
\end{center}\end{figure}

So, for our final example we consider heat transport within a Debye solid ($\kappa\propto (T/T_0)^3,c_p\propto (T/T_0)^3, \rho=\rho_0/(1+\alpha T)\approx\rho_0$, where $T_0$ is the Debye temperature, and $\alpha$ is thermal expansivity ($O(10^{-5}/K$ for a solid)). %(nonlinear transforms of nonlinear media are more numerically stable for diffusion than the wave equation in COMSOL, particularly when no analytic solution exists).
As the nonlinear transform in this case is an arbitrary $g(T)$ that does not satisfy Eq. \ref{eq:Einstein}, we consider a ``phase transition'' transform
\begin{equation}
\lambda(r,T)=\frac{\lambda_L(r)+\lambda_H(r)}{2}+\frac{\lambda_H(r)-\lambda_L(r)}{2}\tanh\frac{T-T_{tr}}{T_\Delta},
\end{equation}
where $\lambda=\rho c_p,\kappa_{rr}$ or $\kappa_{\theta\theta}$, $\lambda_{L(H)}$ are the low (high) temperature transformed parameters, $T_{tr}$ is the transition temperature, and $T_\Delta$ is the range of the intermediate zone.
In particular, we want a cloak for high temperatures (Eq. \ref{eq:TR}, $T>T_{tr}$) and a concentrator for low temperatures.
That is, for $T<T_{tr}$, we have $r^\prime=R_1/R_2*r$ when $0<r<R_2$ and $r^\prime=(R_3-R_1)/(R_3-R_2)*r+(R_1-R_2)/(R_3-R_2)*R_3$ when $R_2<r<R_3$ \cite{TD2}.
COMSOL simulations reveal the nonlinear nature of the steady state, far field temperature distribution (Fig. \ref{fig:Therm}a) and that the low (Fig. \ref{fig:Therm}b) or high (Fig. \ref{fig:Therm}c) temperature cases work as a thermal concentrator or cloak of the Debye solid.
More interesting, when $T(x=0)\equiv T_{tr}\approx8.4T_0$, the device acts like a cloak for $x<0$ and a concentrator for $x>0$, Fig. \ref{fig:Therm}d.

In summary, we have developed a formalism for understanding transformation materials in nonlinear media and undergoing nonlinear transformations, and shown how this formalism can be applied to soliton transport, effective gravitational fields, and thermal management.
It is therefore possible to use nonlinear transformation media to model a wider variety of transport phenomena than have been previously considered.
Furthermore, the constitutive relations that we have derived in Eq. \ref{eq:const} can be used for a wider variety of transformations than just the cloaking transformation.
Given the incorporation of frequency dependent material parameters in the metamaterial realizations of transformation optics \cite{RC1}, the incorporation of nonlinear optical transformations presents the opportunity for a wider variety of functional materials in this framework.
Wave mixing, for example, could be used to generate additional field components at a frequency several times the incident wave's.
If a metamaterial has different resonant effects at these frequencies, it could (say), screen the incident wave while concentrating the nonlinear contribution.
Or it could increase the wave amplitude in a specific region (e.g. via a concentrator), thus increasing the nonlinearity observed in that domain.
Soliton formation could also be promoted, say by counteracting an excessive dispersion or nonlinearity in the external medium.
Generally, nonlinear transformations can increase the versatility of frequency-dependent phenomena in metamaterials and transformation materials.


\begin{thebibliography}{30}
\bibitem{TO1}Leonhardt, U., \textit{Science} \textbf{312}, 1777 (2006).
\bibitem{TO2}Pendry, J.B., Schurig, D. and Smith, D.R., \textit{Science} \textbf{312}, 1780 (2006).
\bibitem{TO3}Leonhardt, U. and Philbin, T.G., \textit{Prog. Opt.} \textbf{53}, 69 (2009).
\bibitem{TO4}Greenleaf, A., Kurylev, Y., Lassas, M., and Uhlmann, G., \textit{SIAM
Rev.} \textbf{51}, 3 (2009).
\bibitem{TO5}Chen, H., \textit{J. Opt. A Pure Appl. Opt.} \textbf{11}, 075102 (2009).
\bibitem{EMCR1}Chen, H., Chan, C. T. and Sheng, P., \textit{Nat. Mater.} \textbf{9}, 387 (2010). %Transformation optics and metamaterials. 387-396
\bibitem{EMCR2}Liu, Y. and Zhang, X., \textit{Nanoscale} \textbf{4}, 5277 (2012). %Recent advances in transformation optics.
\bibitem{EMC1}Shalaev, V. M., \textit{Science} \textbf{322}, 384 (2008).%Transforming Light. 384-386
\bibitem{RC1}Schurig, D. \textit{et al.}, \textit{Science} \textbf{314}, 977-980 (2006).%Metamaterial Electromagnetic Cloak at Microwave Frequencies.
\bibitem{CC1}Li, J.and Pendry, J. B., \textit{Phys. Rev. Lett.} \textbf{101}, 203901 (2008). %Hiding under the Carpet: A New Strategy for Cloaking.
\bibitem{Conc}Yaghjian, A.D. and Maci, S., \textit{New J. Phys.} \textbf{10}(11),  115022, (2008).

\bibitem{TA1}Cummer, S. A. and Schurig, D., \textit{New J. Phys.} \textbf{9}, 45 (2007).%One path to acoustic cloaking.
\bibitem{TA2}Chen, H. and Chan, C. T., \textit{Appl. Phys. Lett.} \textbf{91}, 183518 (2007).%Acoustic cloaking in three dimensions using acoustic metamaterials. 
\bibitem{TA3}Cummer, S. A., \textit{et. al.}  \textit{Phys. Rev. Lett.} \textbf{100}, 024301 (2008).%Scattering Theory Derivation of a 3D Acoustic Cloaking Shell.
\bibitem{TA4}Chen, H. and Chan, C. T., \textit{J. Phys D: Appl. Phys.} \textbf{43}, 113001 (2010).% Acoustic cloaking and transformation acoustics.
\bibitem{TA5}Sklan, S., \textit{Phys. Rev. E} \textbf{81}, 016606 (2010).%Cloaking of the momentum in acoustic waves. 

\bibitem{TD1}Narayana S., and Sato, Y., \textit{Phys. Rev. Lett.} \textbf{108}, 214303 (2012).% Heat Flux Manipulation with Engineered Thermal Materials.
\bibitem{TD2}Guenneau, S., Amra, C., and Veynante, D., \textit{Opt. Express} \textbf{20}, 8207 (2012).% Transformation thermodynamics: cloaking and concentrating heat flux. 8207-8218
\bibitem{TD3}Han, T., \textit{et al.}, \textit{Phys. Rev. Lett.} \textbf{112}, 054302 (2014).%Experimental Demonstration of a Bilayer Thermal Cloak.
\bibitem{TD4}Guenneau, S. and Puvirajesinghe, T. M., \textit{J. R. Soc. Interface} \textbf{10}, 20130106 (2013). % Fick’s second law transformed: one path to cloaking in mass diffusion. 
\bibitem{TD5}Zeng L. and Song, R., \textit{Sci. Rep.} \textbf{3}, 3359 (2013). %Controlling chloride ions diffusion in concrete.
\bibitem{TD6}Schittny, R., Kadic, M., B\"{u}ckmann, T., and Wegener, M., \textit{Science} \textbf{325}, 427 (2014). %Invisibility cloaking in a diffusive light scattering medium.
\bibitem{TD7}Sklan, S.R., Bai, X., Li, B. and Zhang, X., \textit{Sci. Rep.} \textbf{6}, 32915 (2016). %Detecting Thermal Cloaks via Transient Effects.

\bibitem{NEMR}Kadic, M., B\"{u}ckmann, T., Schittny, R., and Wegener, M., \textit{Rep. Prog. Phys.} \textbf{76}, 126501 (2013). %Metamaterials beyond electromagnetism.

\bibitem{PC1}Ruan, Z., Yan, M., Neff, C. W., and Qiu, M., \textit{Phys. Rev. Lett.} \textbf{99}, 113903 (2007). %Ideal Cylindrical Cloak: Perfect but Sensitive to Tiny Perturbations.
\bibitem{PC2}Zhang, B. \textit{et. al.}, \textit{Phys. Rev. B} \textbf{76}, 121101(R) (2007).% Response of a cylindrical invisibility cloak to electromagnetic waves.
\bibitem{PC3}Isi\'{c}, G., Gaji\'{c}, R., Novakovi\'{c}, B., Popovi\'{c}, Z. V., and Hingerl, K.  \textit{Opt. Express} \textbf{16}, 1413 (2008). %Radiation and scattering from imperfect cylindrical electromagnetic cloaks.
\bibitem{PC4}Zolla, F., Guenneau, G., Nicolet, A., and Pendry, J. B., \textit{Opt. Lett.} \textbf{32} 1069 (2007). %Electromagnetic analysis of cylindrical invisibility cloaks and the mirage effect.
\bibitem{PC5}Chen, H., Wu, B.-I., Zhang, B., and Kong, J. A., \textit{Phys. Rev. Lett.} \textbf{99}, 063903 (2007). %Electromagnetic Wave Interactions with a Metamaterial Cloak.

\bibitem{TOGR}Leonhardt, U. and Philbin, T.G., \textit{New J. Phys.} \textbf{8}, 247 (2006).
\bibitem{GRC1}Genov, D.A., Zhang, S., and  Zhang, X., \textit{Nat. Phys.} \textbf{5}, 687 (2009).
\bibitem{GRC2}Chen, H., Miao, R.X. and Li, M., \textit{Optics express}, \textbf{18}, 15183 (2010). %Transformation optics that mimics the system outside a Schwarzschild black hole.
\bibitem{GRC3}Sheng, C., Liu, H., Wang, Y., Zhu, S.N. and Genov, D.A., \textit{Nat. Photonics} \textbf{7}, 902 (2013). %Trapping light by mimicking gravitational lensing. 7(11), 902-906
\bibitem{STC}McCall, M.W., Favaro, A., Kinsler, P., and Boardman, A., \textit{J. Opt.} \textbf{13}, 024003 (2011).

\bibitem{NLTR}Li, Y., Shen, X., Wu, Z., Huang, J., Chen, Y., Ni, Y. and Huang, J.,  \textit{Phys. Rev. Lett.}, \textbf{115}, 195503 (2015). %Temperature-dependent transformation thermotics: From switchable thermal cloaks to macroscopic thermal diodes.
\bibitem{NLTR2}Li, Y., Shen, X., Huang, J. and Ni, Y., \textit{Phys. Lett. A} \textbf{380}, 1641 (2016). %Temperature-dependent transformation thermotics for unsteady states: Switchable concentrator for transient heat flow.

\bibitem{NLO}Boyd, R.W., \textit{Nonlinear Optics (3rd Edition)}, (Academic Press, Orlando, FL 2013).
\bibitem{GR}Misner, C.W., Kip S.T., and Wheeler, J.A., \textit{Gravitation}, (Macmillan, San Francisco, CA 1973).

%\bibitem{Cloak}Schurig, D., \textit{et al.},  \textit{Science}, \textbf{314}(5801), 977-980 (2006). %Metamaterial electromagnetic cloak at microwave frequencies.%Mock, J.J., Justice, B.J., Cummer, S.A., Pendry, J.B., Starr, A.F. and Smith, D.R.
\end{thebibliography}
\end{document}